\documentclass[%
 aip,
 jcp,
 amsmath,amssymb,
 reprint,%
]{revtex4-1}

\usepackage{filecontents}

\usepackage[nohyperlinks, nolist]{acronym}
\usepackage{tikz}
\usetikzlibrary{positioning, fit, shapes.multipart}
\usetikzlibrary{external}
\tikzset{external/force remake}
\usepackage{listings}
\usepackage{fancyvrb}
\usepackage{xr-hyper}
\usepackage{hyperref}
\hypersetup{colorlinks,allcolors=black}
\usepackage{jabbrv}
\usepackage[final]{changes}

\usepackage{graphicx}% Include figure files
\usepackage{dcolumn}% Align table columns on decimal point
\usepackage{bm}% bold math
\usepackage[utf8]{inputenc}
\usepackage[T1]{fontenc}
\usepackage{mathptmx}
\usepackage{etoolbox}

\usepackage{amssymb}
\usepackage{amsmath}

\newcommand{\reffig}[1]{Fig.~\ref{#1}}
\newcommand{\refFig}[1]{Fig.~\ref{#1}} % beginning of sentence
\newcommand{\refeq}[1]{Eq.~\ref{#1}}
\newcommand{\reftab}[1]{Table~\ref{#1}}
\newcommand{\refsec}[1]{Section~\ref{#1}}
\newcommand{\refcite}[1]{Ref.~\onlinecite{#1}}

\usepackage{verbatim}

\makeatletter
\newcommand*{\addFileDependency}[1]{% argument=file name and extension
  \typeout{(#1)}
  \@addtofilelist{#1}
  \IfFileExists{#1}{}{\typeout{No file #1.}}
}
\makeatother

\newcommand*{\myexternaldocument}[1]{%
    \externaldocument[SI-]{#1}%
    \addFileDependency{#1.tex}%
    \addFileDependency{#1.aux}%
}
\myexternaldocument{supplement}

\makeatletter
\def\@email#1#2{%
 \endgroup
 \patchcmd{\titleblock@produce}
  {\frontmatter@RRAPformat}
  {\frontmatter@RRAPformat{\produce@RRAP{*#1\href{mailto:#2}{#2}}}\frontmatter@RRAPformat}
  {}{}
}%
\makeatother
\begin{document}

%TC:ignore
\preprint{AIP/123-QED}

\title[ELECTRODE]{ELECTRODE: An electrochemistry package for atomistic simulations}
\author{Ludwig J. V. Ahrens-Iwers}
\affiliation{Institute of Advanced Ceramics, Hamburg University of Technology, Hamburg, Germany}%Lines break automatically or can be forced with \\

\author{Mathijs Janssen}%
\affiliation{Department of Mathematics, Mechanics Division, University of Oslo, N-0851 Oslo, Norway}%

\author{Shern R. Tee}%
\affiliation{Australian Institute for Bioengineering and Nanotechnology, The University of Queensland, Brisbane, Queensland, Australia}%

\author{Robert H. Mei{\ss}ner}
\affiliation{Institute of Polymers and Composites, Hamburg University of Technology, Hamburg, Germany}%
\affiliation{Helmholtz-Zentrum Hereon, Institute of Surface Science, Geesthacht, Germany}

\email{s.tee@uq.edu.au, robert.meissner@tuhh.de}

\date{\today}% It is always \today, today,

\begin{abstract}
Constant potential methods (CPM) enable computationally efficient simulations of the solid-liquid interface at conducting electrodes in molecular dynamics (MD).
They have been successfully used, for example, to realistically model the behavior of ionic liquids or water-in-salt electrolytes in supercapacitors and batteries.
The CPM models conductive electrodes by updating charges of individual electrode atoms according to the applied electric potential and the (time-dependent) local electrolyte structure.
Here we present a feature-rich CPM implementation, called ELECTRODE, for the Large-scale Atomic/Molecular Massively Parallel Simulator (LAMMPS), which includes a constrained charge method and a thermo-potentiostat.
The ELECTRODE package also contains a finite-field approach, multiple corrections for non-periodic boundary conditions of the particle-particle particle-mesh solver, and a Thomas-Fermi model for using non-ideal metals as electrodes.
We demonstrate the capabilities of this implementation for a parallel-plate electrical double-layer capacitor, for which we have investigated the charging times with the different implemented methods and found an interesting relationship between water and ionic dipole relaxations.
To prove the validity of the one-dimensional correction for the long-range electrostatics, we estimated the vacuum capacitance of two co-axial carbon nanotubes and compared it to structureless cylinders, for which an analytical expression exists.
In summary, the ELECTRODE package enables efficient electrochemical simulations using state-of-the-art methods, allowing one to simulate even heterogeneous electrodes.
Moreover, it allows unveiling more rigorously how electrode curvature affects the capacitance with the one-dimensional correction.
\end{abstract}
%TC:endignore

\acresetall

\maketitle

\section{Introduction}
\label{sec:intro}

A common approach to treating electrodes in atomistic simulations is to assume them to be uniformly charged walls, either structureless or atomically resolved. 
In the case of equilibrium electrolytes near planar electrodes at low charge densities, this approach is known to capture the electrochemical properties well.
Several studies, however, have emphasized the importance of polarization of the electrodes by the ions and molecules in their vicinity \cite{Merlet2013,Breitsprecher2015ElectrodeCapacitors,Vatamanu2018ApplicationFields,Haskins2016EvaluationLayers,Gading2022ImpactLiquids}. 
In more realistic electrochemical scenarios, \ac{cpm} \ac{md} results are often significantly different from those obtained with uniformly charged electrodes\cite{Merlet2013b}.

While the \ac{cpm}\cite{Siepmann1995InfluenceSystems, Reed2007ElectrochemicalElectrode} is a popular tool for modeling metal electrodes by dynamically updating individual charges on electrode atoms, alternatives such as image charge methods \cite{Tyagi2010, Petersen2012, Dwelle2019ConstantSimulation, Nguyen2019} are commonly used to enforce a constant potential for planar electrodes.
While one of these methods can handle non-planar surface by inducing a charge density on the interface between two media most of them are limited to planar electrodes.\cite{Nguyen2019}
Those approaches faithfully reproduce the behavior of electrolytes near electrodes, particularly the correlation between thermal fluctuations in the electrolyte near the electrode and the induced-charge polarization of the electrode, while obtaining a realistic picture of the electrical double-layer.
A \ac{cpm} \ac{md} is able to capture the temporal response in the build-up and break-down of electric double-layer and thereby allows realistic capacitor charging and discharging curves to be generated \emph{in silico} \cite{Merlet2013b,Breitsprecher2020HowNanopores,Kondrat2014AcceleratingPores,Kondrat2016PressingSupercapacitors}.
Interestingly, near highly charged planar electrodes \cite{Wang2014EvaluationCapacitors, Limmer2013ChargeCapacitors, Merlet2014TheOwn} and non-planar electrodes (such as curved substrates or nanoporous carbons) \cite{Merlet2012, Merlet2013b, Merlet2013HighlySupercapacitors, Lahrar2021Carbon-carbonCapacitance, Forse2016NewSupercapacitors, Seebeck2020a, Seebeck2022ElucidatingSupercapacitors, McDaniel2022CapacitanceSimulations}, \ac{cpm} \ac{md} and Monte Carlo simulations \cite{Caetano2021MonteMacroions} yield a spatially-specific charge polarization and a non-trivial electrolyte structure. 

Here, we present a package for treating electrodes in \ac{md} simulations which interfaces with \ac{lammps} \cite{Thompson2022LAMMPSScales}.
Our ELECTRODE package uses the highly parallelized and efficient computational infrastructure of \ac{lammps} and allows interaction with many other packages and features already available in \ac{lammps}.
This work builds in part on an earlier work in which we showed how a \ac{p3m}-based calculation makes the electrostatic calculations of a \ac{cpm} simulation more efficient\cite{Ahrens-Iwers2021}.
In addition to some new enhancements to the \ac{cpm}, this implementation provides a \ac{ccm} and a \ac{tp}\cite{Deienbeck2021DielectricApproach}.
To capture the electronic response of non-ideal metals, a \ac{tf} model\cite{Scalfi2020ASimulations} is included.
Both an Ewald and a \ac{p3m} $\mathbf k$-space solver are available for various constraints of the periodicity of the systems, such as infinite slabs, cylinders or fully 3D periodic systems. 
The ELECTRODE package also contains the closely-related \ac{ff} \cite{Dufils2019SimulatingElectrode} and \ac{fd} \cite{Dufils2021ComputationalSimulations} methods, which extend \ac{cpm} \ac{md} with a slab geometry to fully periodic boundary conditions for increased computational efficiency.\cite{Tee_2022}

A list of new features in the ELECTRODE package is presented in \refsec{sec:features} including a brief description of their theoretical background.
In \refsec{sec:results}, we summarize the concept of \ac{cpm} \ac{md} using data from various \ac{cpm} \ac{md} runs and rationalize it based on the charging times of an electrical double-layer capacitor.
We discuss briefly for which situation each approach is suitable and give in \refsec{sec:conclusions} an overview of future development directions and possible applications. 

\section{Features\label{sec:features}}
% TODO  p3m (including matrix settings), hybrid pair-style, custom neighbor lists,  variable inputs

\subsection{Constant potential method}
\label{sec:feat-cpm}

In atomistic electrochemical simulations, the system of interest is often a fluid electrolyte confined between two electrodes (cf. inset of \reffig{fig:tau}).
This could serve as an \emph{in silico} nanoscale model of a capacitor, to optimize some metric, such as energy or power density, by modifying electrolyte composition or electrode structure.

The distinctive feature of \ac{cpm} \ac{md} is the calculation of electrode charges which keep electrodes at a desired electrostatic potential.
To achieve this, we first partition the potential energy, $U$, of an MD simulation into:
\begin{equation}
    U = U_{\mathrm{non-Coul}} + U_{\mathrm{elyt}} + U_{\mathrm{elec}}
    \label{eqn:u_partition}
\end{equation}
Here $U_{\mathrm{non-Coul}}$ includes all non-Coulombic interactions, $U_{\mathrm{elyt}}$ includes all Coulombic interactions between electrolyte particles, and $U_{\mathrm{elec}}$ includes all Coulombic interactions involving electrode particles (both with electrolyte particles, and with other electrode particles).
While the former are treated with regular force field approaches, the last term is treated somewhat special.
$U_{\mathrm{elec}}$, i.e., without electrolyte-electrolyte interactions, is written in terms of an electrode charge vector $\mathbf{q}$ comprising all electrode charges as
\begin{equation}
    U_{\mathrm{elec}}\left(\{\mathbf r\}, \mathbf q\right) = \frac{1}{2} \mathbf q^{\mathrm T} \mathbf A \mathbf q - \mathbf b^{\mathrm T} (\{\mathbf r\}) \mathbf q - \mathbf v^{\mathrm T} \mathbf q
    \label{eqn:coulomb}
\end{equation}
with a matrix $\mathbf{A}$, and vectors $\mathbf{b}$ and $\mathbf{v}$; where $\mathbf{b}$ depends on the electrolyte positions $\{\mathbf{r}\}$.
The applied potential $\mathbf v$ has an entry for every electrode atom.
The interactions between electrode atoms are represented by $\mathbf A$, called elastance matrix due to the analogy between a vacuum capacitor and a spring.
If the electrode atoms do not move, $\mathbf{A}$ can be pre-computed, allowing significant computational savings. 
The electrolyte vector $\mathbf{b}(\{\mathbf{r}\})$ represents the electrostatic potential on each electrode atom due to the electrolyte atoms.

At each step, $\mathbf{q}$ is updated to minimize the Coulombic energy contribution $U_\mathrm{elec}$, possibly subject to additional constraints. The desired energy-minimizing charge vector $\mathbf{q}^*$ is straightforward to calculate\cite{Scalfi2020}:
\begin{equation}
    \mathbf q^* = \mathbf A^{-1} \left[\mathbf b(\{\mathbf r\}) +\mathbf v \right].
    \label{eqn:cpm}
\end{equation}
Here, the elastance has been inverted to yield $\mathbf A^{-1}$, which is called the capacitance matrix in light of its role in equation \eqref{eqn:cpm}:
the response of the charge vector $\mathbf{q}^*$ can calculated as the product of the capacitance matrix with the vector of external potentials, analogously to the well-known scalar equation $Q = CV$ linking the capacitance $C$ to the charge $Q$.

Provided the electrode atom positions and thus the vacuum capacitance are constant, the main computational burden is the calculation of $\mathbf b(\{\mathbf r\})$ at every time step, which is necessary due to the motion of the electrolyte.
The primary purpose of ELECTRODE is to compute the electrode-electrolyte interaction in $\mathbf b$ efficiently and update the electrode charges accordingly.
Alternatively, the electrode charges could be obtained with the conjugate gradient method, which solves the minimization problem without a matrix inversion.\cite{Vatamanu2010MDSim,Li2021}
Yet another approach is to treat the electrode charges as additional coordinates and perform mass-zero constrained dynamics for them\cite{Coretti2020Mass-zeroSystems}.

In \ac{md} with periodic boundary conditions, the simulation cell ideally is charge neutral. 
\citet{Scalfi2020} showed that this constraint could be imposed by using the symmetric matrix
\begin{equation}
    \label{eqn:S}
    \mathbf S \equiv \mathbf A^{-1} - \frac{\mathbf A^{-1} \mathbf e \mathbf e^ \mathrm T \mathbf A^{-1}}{\mathbf e^\mathrm T \mathbf A^{-1} \mathbf e}\; , \quad \mathbf e^\mathrm T = (1, \dots, 1)
\end{equation}
as capacitance matrix instead of $\mathbf A^{-1}$.

Non-ideal metallic electrodes have been recently modeled by \citet{Scalfi2020ASimulations} using a semiclassical \ac{tf} approach.
We have implemented this promising approach in our ELECTRODE package, as its implementation is very similar to the self-interaction correction of the Ewald summation\cite{Hu2014InfiniteInterfaces} and contains only a single summation over the electrode atoms.
An interesting alternative to effectively model a wide range of materials between insulator and ideal metal was proposed by \citet{Schlaich2022ElectronicSurfaces} and involved using a virtual \ac{tf} fluid within the electrodes.
However, the virtual \ac{tf} fluid approach appears computationally more expensive.
Both models require free parameters, most crucially the \ac{tf} length, $l_\mathrm{TF}$, in \citet{Scalfi2020ASimulations} and (a rather artificial) parameterization of the virtual \ac{tf} fluid in the approach of \citet{Schlaich2022ElectronicSurfaces}.
Further, assumptions such as atom-centered densities prohibit effects such as quantum spillover and delocalization of the image plane.
It should be noted that ELECTRODE provides a more flexible implementation of the \ac{tf} model, allowing heterogeneous electrodes with different $l_\mathrm{TF}$ for different atom types.
For more general information on \ac{cpm} \ac{md} approaches, the interested reader is referred to the excellent and thorough review of current electrode-electrolyte simulations by \citet{Scalfi2020MolecularInterfaces}, the well-written theory part of \textit{MetalWalls}\cite{mw-docu} or the thesis of \citet{Gingrich2010}.

\subsection{Simulating an arbitrary number of electrodes}
\label{sec:feat-n-electrodes}

A \ac{cpm} \ac{md} is typically performed with two electrodes, which means there are only two possible values for each of the $n$ components of the potential in \refeq{eqn:coulomb}. %, e.g. $v_i \in [v_\mathrm{top}, v_\mathrm{bot}]$ in case of a simple plate capacitor.
In the ELECTRODE package, an arbitrary number $N$ of electrodes is allowed with every electrode atom belonging to exactly one electrode.
We define an electrode-wise indicator vector $\mathbf g_\alpha$ for every electrode $\alpha$ with $n$ entries, which are equal to $1$ if the respective electrode particle belongs to that electrode and $0$ otherwise.
The indicator matrix
\begin{equation}
    \mathbf G = \begin{bmatrix} \mathbf g_1 & \mathbf g_2 &  \cdots & \mathbf g_N \end{bmatrix}
\end{equation}
comprising the indicator vectors of all $N$ electrodes allows us to connect electrode-wise quantities to particle-wise quantities.
From hereon, we use tildes for electrode-wise quantities.
For instance, we define $\mathbf{\tilde v}$ as the electrode-wise potential and use it to write the potential $\mathbf v = \mathbf G \mathbf{\tilde v}$.
%In the following a tilde will be used for electrode-wise quantities.
%
Likewise, energy-minimizing charges $\mathbf{q^*}$ for a given set of electrode-potentials are
\begin{equation} \mathbf{q^*} = \mathbf{S} (\mathbf{b} + \mathbf{v}) = \mathbf{S} \mathbf{b} + \mathbf{S} \mathbf{G} \mathbf{\tilde{v}}.
\label{eqn:electrode-wise}
\end{equation}
%
%In ELECTRODE, we calculate and cache the ``electrode response vector array'' $\mathbf{SG}$, an $n$-by-$N$ matrix, and store the imposed potentials as a $N$-sized vector $\mathbf{\tilde{v}}$.
%This scheme can generate some computational savings, in that time-varying potentials require only a smaller per-electrode, $N$-sized vector to be updated each time step. 
%But more importantly the resulting code is tidier and conceptually simpler, and can describe an arbitrary number of electrodes simply by changing the shapes and sizes of $\mathbf{SG}$ and $\mathbf{\tilde{v}}$.

\subsection{Simulating electrodes at specified total charge}
\label{sec:feat-ccm}

In the \ac{ccm}, the user sets the electrode-wise total charge $\mathbf{\tilde{q}^*}$ for each electrode. Such a fixed-charge setup corresponds to an open-circuit configuration\cite{Jeanmairet2022MicroscopicCapacitors}. 
This type of simulation has recently been attempted as a variation of the finite-field method\cite{Dufils2021ComputationalSimulations}.
Therein, it was found that ramping the total charge up or down over time could be considered computational amperometry, and a faster non-equilibrium response was observed.

Working with the capacitance matrix $\mathbf{A}^{-1}$ rather than the symmetrized matrix $\mathbf{S}$ (since charge neutrality is explicitly enforced by the appropriate choice of $\mathbf{\tilde{q}^*}$), we have:
\begin{equation}
     \mathbf{\tilde{q}^*} = \mathbf{G}^\mathrm{T} \mathbf{q^*} = \mathbf{G}^\mathrm{T} \mathbf{A}^{-1} \mathbf{b} + \mathbf{G}^\mathrm{T} \mathbf{A}^{-1} \mathbf{G} \mathbf{\tilde{v}} 
     \equiv \mathbf{\tilde{q}_b^*} + \mathbf{\tilde{C}} \mathbf{\tilde{v}}.
     \label{eqn:ewise-conq}
\end{equation}
$\mathbf{\tilde{q}_b^*}$ defines the total charge each electrode would carry at zero potential, and $\mathbf{\tilde{C}}$ is the electrode-wise capacitance matrix.
To subsequently estimate $\mathbf{\tilde{q}_b^*}$, \refeq{eqn:ewise-conq} is solved for $\mathbf{\tilde v}$ which is then applied using the \ac{cpm}.
This results in an energy minimization w.r.t. the charge distribution with a constraint on the total electrode charges.
Analogous to how constant volume and constant pressure simulations can be thermodynamically equivalent, \ac{ccm} and \ac{cpm} simulations will give the same capacitances under suitable conditions.
However, a thorough proof of that assertion is out of scope of this work and will be discussed in an upcoming work.
\subsection{Simulating electrodes with a thermo-potentiostat}
\label{sec:feat-tp}
\citet{Deienbeck2021DielectricApproach} recently presented a \acf{tp} that takes into account the fluctuation-dissipation relation of electrode charges at a given voltage and temperature in an electronic circuit.
They have also provided a \ac{tp} implementation based on a uniform charge distribution using the scripting capability of \ac{lammps}\cite{Thompson2022LAMMPSScales}.
The ELECTRODE package provides an implementation that minimizes the energy with respect to the charge distribution and conforms to the formalism described by \citet{Deienbeck2021DielectricApproach}.
Our \ac{tp} approach is currently limited to only two electrodes and instead of a vector of applied potentials $\mathbf{\tilde v}$, a potential difference 
\begin{equation}
\Delta v_0 = v_\mathrm{top} - v_\mathrm{bot}
\label{eqn:delta-v}
\end{equation}
between two electrodes is used.
At every time step, the potential difference $\Delta v(t)$ between the two electrodes is evaluated to find the new capacitor charge according to
\begin{align}
    q(t + \Delta t) = &q(t) - C_0\left[\Delta v(t) - \Delta v_0\right]\left(1 - \mathrm e ^{-\Delta t / \tau_v} \right) \nonumber \\
        &+ X \sqrt{k_\mathrm{B} T_v C_0 \left(1 - \mathrm e ^{-2 \Delta t / \tau_v} \right)}.
    \label{eqn:deissenbeck}
\end{align}
Here, $k_\mathrm{B}$ is the Boltzmann constant, $\tau_v$ and $T_v$ are parameters of the \ac{tp} and $X$ is a normally distributed random number with a mean of 0 and a standard deviation of 1.
The vacuum capacitance $C_0$ is obtained from the capacitance matrix\cite{Scalfi2020} and the effective potential $\Delta v(t)$ is computed from the electrode charges and the electrolyte configuration (cf. \refeq{eqn:ewise-conq}).
Hence, all quantities required to evaluate \refeq{eqn:deissenbeck} are readily available in the \ac{cpm}.
The obtained capacitor charge $\pm q(t+\Delta t)$ is applied using the \ac{ccm} on both electrodes, respectively. % protocol outlined in \refsec{sec:feat-ccm}.

\subsection{Simulations with different periodicity}
\label{feat-boundcorr}
The Ewald summation commonly assumes periodic boundary conditions in all three directions and has to be modified for systems with slab and one-dimensional periodic geometries.
As shown by \citet{Smith1981ElectrostaticCrystals}, a regular 3D Ewald summation for slab-like systems, which are periodic in the $xy$-plane but confined in $z$-direction, results in a dipole term 
\begin{equation}
    J^{\text{2D}}(\mathbf M)=\frac{2\pi}{V}M_z^2.
    \label{eqn:ew3dc_slab}
\end{equation} 
$M_z$ is the $z$-component of the dipole of the simulation cell.
This dipole term was subsequently used for correcting the infinite boundary artifact of slab-like systems\cite{Yeh1999}.
This is known as the \acs{ew3d}c method, which is implemented in many \ac{md} codes including \ac{lammps}\cite{Thompson2022LAMMPSScales}.
%
%Alternatively, recently \citet{Hu2014InfiniteInterfaces} and earlier also others\cite{DeLeeuw1979ElectrostaticLattices,Heyes1977MolecularChloride,Parry1975TheCrystal} 
Several authors \cite{Hu2014InfiniteInterfaces,DeLeeuw1979ElectrostaticLattices,Heyes1977MolecularChloride,Parry1975TheCrystal} have shown that the infinite boundary contribution in slab-like geometries can also be solved in an exact form. 
This rarely implemented  EW2D solver is another cornerstone of the ELECTRODE package.

Just like slab-like geometries, systems with only one periodic dimension require an appropriate treatment of the long-range electrostatic interactions.
As shown by \citet{Brodka2004Three-dimensionalDirection}, the approach of \citet{Smith1981ElectrostaticCrystals} can be extended for an infinitely extended one-dimensional summation:
\begin{equation}
    J^{\text{1D}}(\mathbf M)=\frac{\pi}{V}\left(M_x^2 + M_y^2\right).
    \label{eqn:ew3dc_wire}
\end{equation}
Here, $z$ is the periodic dimension and $M_x$ and $M_y$ are the respective components of the total dipole of the unit cell.
Contrasting established codes, the ELECTRODE package contains these corrections for one-dimensional periodic systems. 
And even more crucially, ELECTRODE is the first package (as far as we know) to implement these corrections in combination with a CPM.
As an outlook, the electrostatic layer correction\cite{Arnold_2002} in combination with PPPM\cite{de_Joannis_2002} is also considered for implementation as an alternative to the \acs{ew3d}c approach.

As demonstrated exemplarily for slab-like two-dimensional periodic systems, the boundary corrections can be easily incorporated into the \ac{cpm} formalism by splitting the dipole components into their electrode and electrolyte contributions:
\begin{align}
    J^{\text{2D}} 
    &= \frac{2 \pi}{V}\left[(M_z^\text{elec})^2 + 2 M_z^\text{elec}M_z^\text{elyt} + (M_z^\text{elyt})^2\right]\nonumber \\
    &= \frac{2 \pi}{V}\left[\sum_{ij}z_i z_j q_i q_j + 2 M_z^\text{elyt}\sum_i z_i q_i +(M_z^\text{elyt})^2 \right].
\end{align}
This way dipole corrections fit into the linear form of the Coulombic energy in \refeq{eqn:coulomb} that is used in the \ac{cpm} and the computational effort for the electrode-electrolyte interaction scales linearly with the number of particles.

\subsection{Simulating electrodes with the \acl{ff} method}\label{chp:finite_field}

In the \ac{ff} method \cite{Dufils2019SimulatingElectrode}, the potential difference between two electrodes is not directly specified using the applied potential $\mathbf{v}$. 
Instead, the simulation cell is periodic in the $z$ direction, i.e., without adding the artificial vacuum between the slabs required otherwise. 
The \ac{ff} method allows efficient simulations of infinite electrode slabs, since no additional vacuum is required.
However, complexly shaped electrodes or electrodes with a one-dimensional periodicity cannot be simulated with the \ac{ff} method.

A potential difference $\Delta v_0$ (cf. \refeq{eqn:delta-v}), is created in the \ac{ff} method by introducing a $z$-directed electric (polarization) field of magnitude $-\Delta v_0 / L_z$, creating a discontinuity of $\Delta v_0$ across the periodic $z$ boundary (and thus between the two electrodes on either side of the slab).
In this formulation, the electrode Coulombic energy is
\begin{equation}
    U_\mathrm{elec} = \frac{1}{2}\mathbf{q}^\mathrm{T}  \mathbf{A} \mathbf{q} - \mathbf{b}^\mathrm{T}(\{ \mathbf{r}\}) \mathbf{q} + \Delta v_0 \pmb \zeta^\mathrm{T} \mathbf{q}.
\end{equation}
Here, $\pmb{\zeta}$ is a vector containing the normalized $z$-positions of each electrode atom, namely $\{\mathbf{z}/L_z\}$ with an offset for the bottom electrode to make the system symmetric along the $z$-direction, replicating the conductor-centered supercell in \refcite{Dufils2019SimulatingElectrode}. 
The energy-minimizing charge $\mathbf{q}^*$ in this model is
\begin{equation}
    \mathbf{q}^* = \mathbf{S}[\mathbf{b}(\{ \mathbf{r} \}) - \Delta v_0 \pmb \zeta]
    \label{eqn:cpm-ffield}
\end{equation}
which is equivalent to the standard \ac{cpm},
replacing $\mathbf{v}$ with $-\Delta v_0 \pmb \zeta$.

\section{Results and discussion}
\label{sec:results}

\subsection{Charging times}

A simple capacitor model is adapted from an example in the \textit{MetalWalls} repository\cite{Marin-Lafleche2020} and comprises a saline solution between two gold electrodes with three layers each.
To compare the equilibrium conditions of the \ac{cpm} and \ac{ccm} we calculated the capacitance per area from the averaged equilibrium charges and voltages at an applied voltage of 2\,V and charge of 4.4\,$e$, respectively.
The obtained values of $2.94$\,{\textmu}F\,cm$^{-2}$ and $2.91$\,{\textmu}F\,cm$^{-2}$ for the \ac{cpm} and \ac{ccm} respectively differ by only 1\,\%, showing a good agreement between the methods at equilibrium.
%

\begin{figure}
    \centering
    \includegraphics{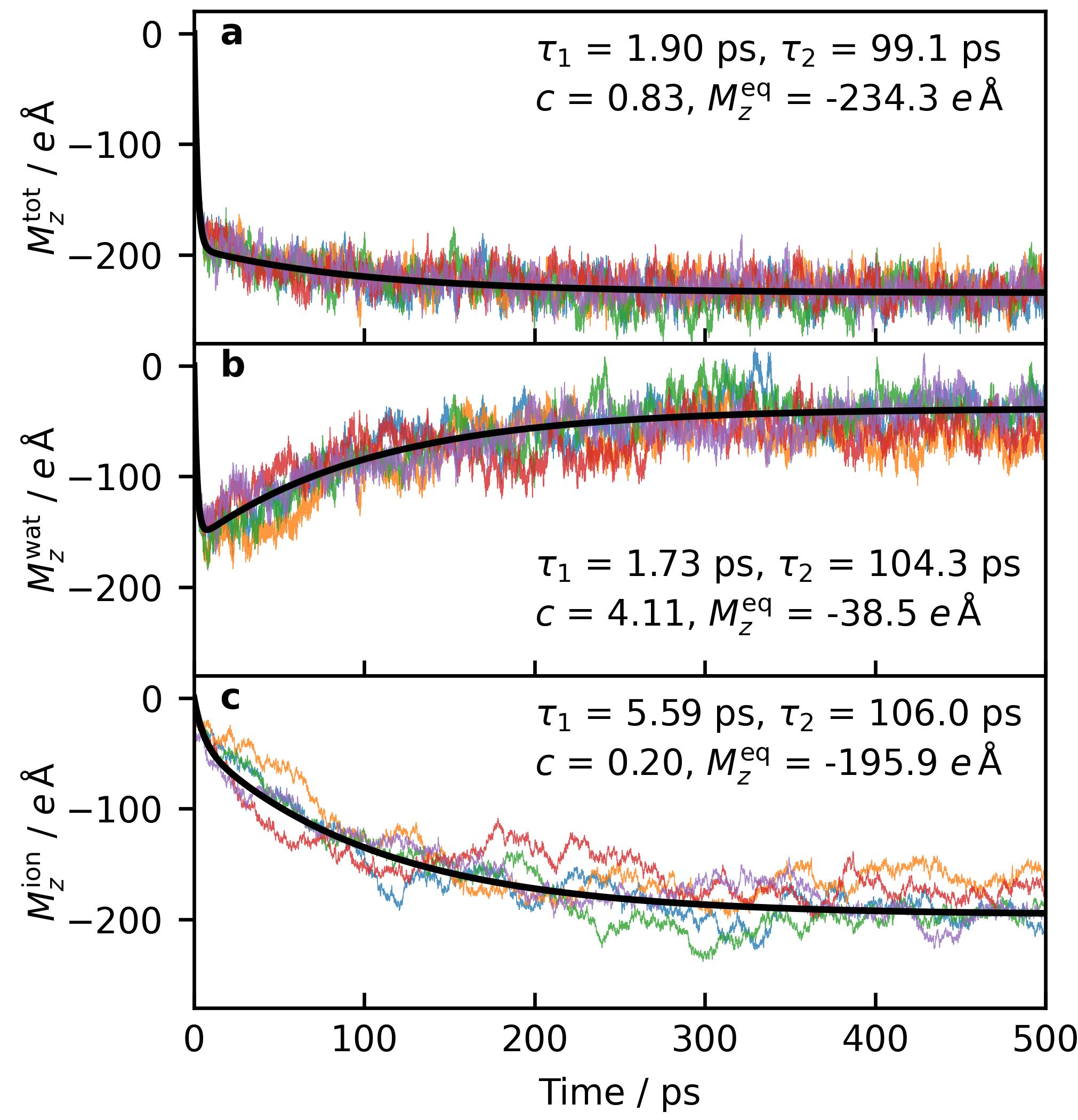}
    \caption{%
            Individual fitting parameters of the bi-exponential function (cf. \refeq{eqn:dipole_fit}) applied to: (a) the total dipole $M_z^\text{tot}$, (b) the dipole of water molecules $M_z^\text{wat}$ and (c) to the dipole of ions in the electrolyte  $M_z^\text{ion}$ from a standard \ac{cpm} simulation at 2\,V.
            While only five of the 100 trajectories are shown for illustration, fits are made to the entire set of trajectories.
            }
    \label{fig:separate-tau}
\end{figure}

%In contrast to a \ac{ccm}, 
When a voltage is initially applied with the \ac{cpm}, the charge induced on both electrodes is very small since the capacitance of the electrode pair \emph{in vacuo} is small \cite{Scalfi2020, Ahrens-Iwers2021}. 
However, the electrode charges induce the formation of a dipole in the electrolyte, which in turn induces additional charge on the electrode.
\ac{cpm} \ac{md} thus models the process of charging an electrical double layer capacitor physically correctly, and the charging-discharging curves obtained from \ac{cpm} \ac{md} can be used to fit parameters for equivalent macroscopic electrical circuits \cite{Sampaio2020ComparingSimulations}.
Two charging times $\tau_1$ and $\tau_2$ are obtained by fitting a bi-exponential charging function 
\begin{equation}
    M_z(t) = M_z^\mathrm{eq} \left[1 - c \exp(-t / \tau_1) - (1-c)\exp(-t / \tau_2)\right]
    \label{eqn:dipole_fit}
\end{equation} 
to the $z$-component of the electrolyte dipole\cite{Noh2019UnderstandingSimulations}.
$M_z^\mathrm{eq}$ is the extrapolated equilibrium dipole reached at late times.
Comparing the individual contributions to the total electrolytic dipole reveals that $\tau_1$ describes relatively fast water dipole relaxations and $\tau_2$ describes charging times related to the ion diffusion.
To validate this statement, we show both individual components in \reffig{fig:separate-tau}.
From the mixing parameters $c$ in the panels, it is clear that water dipole relaxation dominates at the beginning, while ion diffusion prevails at later times.
$\tau_1$ of the water in \reffig{fig:separate-tau}b largely corresponds to that of the total dipole at the beginning of the charging.
While the ions in \reffig{fig:separate-tau}c relax mainly on the slower timescale of $\tau_2$, the water dipole decreases as the ion dipole is slowly built up. 

\begin{figure}
    \centering
    \includegraphics{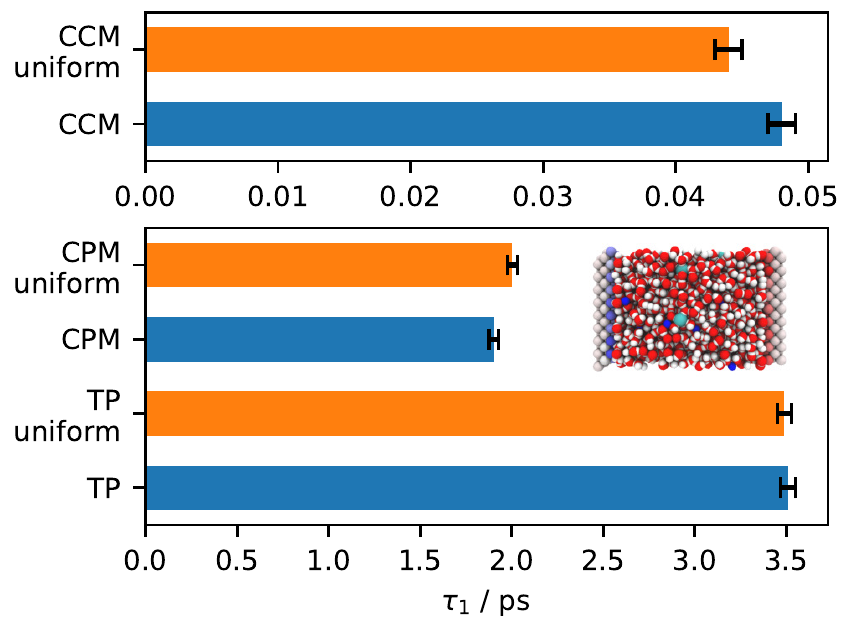}
    \caption{%
            Charging time constants $\tau_1$ and empirical standard deviations employing different \ac{cpm} approaches as denoted on the $y$-axis.
            Inset shows the model system used to compare the different approaches, consisting of an aqueous NaCl electrolyte in contact with two gold surfaces.
            Note the different charging time constants $\tau_1$ for the \ac{ccm} and \ac{cpm} approaches.
            }
    \label{fig:tau}
\end{figure}

\refFig{fig:tau} compares results using \ac{cpm}, \ac{ccm} and \ac{tp}, the latter with a time constant $\tau_v = 100$\,fs, to each other and to their uniformly-charged counterparts.
We focus on the charging term with the faster timescale $\tau_1$ because of its large contribution to the total dipole.
For the uniform variants, the charges are always evenly distributed across the inner layers of the electrodes.
The uniform methods are in general very close to their heterogeneous counterparts, which is consistent with previous studies that found only small differences between a heterogeneous and uniform \ac{cpm} at low voltages for simple planar electrodes \cite{Wang2014EvaluationCapacitors}.
%
%
% method M^0
% cpm    -211
% cpm tf -169
%
The charging times obtained with \ac{ccm} appear to be too fast, since $\tau_1$ is about two orders of magnitude smaller than obtained with the potential-based methods, which is a well-known effect \cite{Vatamanu2011, Merlet2013b} that could be related to rather unphysical high temperatures and voltages when applying an instantaneous charge with \ac{ccm} on the electrodes\cite{Merlet2013b}.
\ac{ccm} also predicts two times smaller $\tau_2$ than other methods, while the contribution of the second exponential decay is almost negligible since $c=0.98$ in \refeq{eqn:dipole_fit}.
However, these values should be interpreted with caution, given that the bi-exponential curve is a poor fit in the case of a \ac{ccm} (cf. \reffig{SI-fig:bi-exp}).
Using the \ac{tp} little change in the slower timescale $\tau_2$ is observed compared to the \ac{cpm} and it only weakly depends on the time constant $\tau_v$ (cf. \reftab{SI-tab:fits}).
In contrast, the applied voltage $v_0$ in a \ac{cpm} has an effect on the ionic charging times $\tau_2$ and for a smaller applied voltage of 0.3\,V, $\tau_2$ drops to 50\,ps (cf.~\reffig{SI-fig:voltage-tau}).

Using a \ac{tf} model for representing real metals impacts both charging times and the total dipole.
It is interesting to note that while a decrease for $\tau_1$ with increasing $l_\mathrm{TF}$ in \reffig{SI-fig:tau-tf-tps}a is observed, interpretation of $\tau_2$ is more complex when using a \ac{tf} (cf. \reftab{SI-tab:fits}). 
Especially when comparing this to a regular \ac{cpm}, i.e. $l_\mathrm{TF}=0$, at the same voltage. 
The quite significant difference between a regular \ac{cpm} and the \ac{tf} for small $l_\mathrm{TF}$ might be an artifact due to the rather thin metal slab model or is due to a complex interplay between charge screening in the metal and the water/ionic relaxation and/or the smaller total dipole obtained with \ac{tf} model.

To understand the range of the second relaxation time $\tau_2\approx50-100$\,ps, it is instructive to consider the product $RC$ of the aforementioned areal capacitance $C=2.94$\,{\textmu}F\,cm$^{-2}$ and the areal electrolyte resistance $R$. % ionic relaxation time 
Continuum models for ion dynamics have shown that the ionic relaxation time decently agrees with $RC$ for applied potentials up to around the thermal voltage $e/(k_\mathrm{B} T)\approx25$\,mV;\cite{Bazant2004diffuse,Janssen2019CurvatureElectrodes}
%$\tau_{\rm ion}$ also follows from the  equations for cases where the .
%
a recent dynamical density functional theory for a dense electrolyte found that the ions relaxed with $RC$ even around 1\,V.\cite{Ma2022dynamics}
In a bulk electrolyte at infinite dilution, the areal resistance between two electrodes spaced $L$ apart is given by $R=L\varrho$, where $\varrho=k_\mathrm{B} T/(2e^2 D c_\mathrm{b})$ is the ionic resistivity, $k_\mathrm{B} T$ is the thermal energy, $e$ is the elementary charge, $D$ is the ionic diffusion constant, and $c_\mathrm{b}$ is the salt number density.\cite{Avni_2022}
In our simulations, the plate separation was $L=5$\,nm and the salinity in the bulk phase was approximately 0.95\,M, corresponding to $c_\mathrm{b}\approx0.57$\,nm$^{-3}$; the ion diffusivity $D\approx1.5\cdot10^{-9}$\,m\textsuperscript{2}/s was obtained from a separate bulk electrolyte simulation.
Using these values, we obtained $RC=14$\,ps; roughly 7 times smaller than the largest fitted $\tau_2$.
This discrepancy must be due partly to our underestimation of $\varrho$, which, at the salinity of our interest, is larger by a factor of about 1.7;\cite{Avni_2022} accounting for this effect yields a relaxation time of $RC=23$\,ps.
Another cause of the remaining factor 4 discrepancy between the largest fitted $\tau_2$ and predicted ionic relaxation times is the nanoconfinement, which could affect the diffusivity $D$ and, in turn, the areal resistivity $R$.
Finally, the mentioned increase of $\tau_2$ with the applied potential is in line with the potential dependence of the capacitance of the Gouy-Chapman model, though in disagreement with that of the Kilic-Bazant-Ajdari\cite{Kilic_2007} model.  
%From the point of view of an analytical model \cite{Bazant2004diffuse}, this increase in ionic charging times with increasing potential is not surprising, although one would expect charging times to be orders of magnitude slower with increasing voltage.
%
Although the analytical estimates of $RC$ times presented here are interesting and provide starting points for further research on the implications of nanoconfinement and finite salt concentration on charging times, these results should not be overinterpreted as the analytical models contain simplifications that may not apply to such nanoscopic systems.

%\begin{align}\label{eq:RC}
%    \tau_{\rm ion}=\frac{\lambda_\mathrm{D} L}{2D(1+r_{\rm ion}/\lambda_\mathrm{D})}\,,    
%\end{align}
%
%, $r_{\rm ion}$ is the ion radius, and $\lambda_\mathrm{D}=\sqrt{\varepsilon_0 \varepsilon_r k_\mathrm{B} T/(2e^2 c_b)}$ is the Debye length,, and $\varepsilon_0$ and $\varepsilon_r$ are the vacuum and relative permittivity, respectively.
%
%$\tau_{\rm ion}$ is an estimate for the ionic $RC$ time, which follows from multiplying an electrolyte's resistance with the electric double layer capacitance, estimated here as a harmonic average of the Helmholtz and Stern layer capacitance\cite{Ma2022dynamics}.
%
%
%
%\citet{Deienbeck2021DielectricApproach} showed that the permittivity of nanoconfined water was substantially smaller than in bulk. 
%
%Indeed, by determining the static permittivity from the relation $\epsilon_r= C/C_0=16.8$ with the capacitance $C = \langle Q\rangle / v_0$ calculated from the average charge of the last 100\,ps of the \ac{cpm} trajectories
%and  $\tau_{\rm ion}\approx100$\,ps, which is in good agreement with our fitted $\tau_2$.
%
%\textcolor{red}{The \ac{tp} could potentially be used to model a resistor with resistance $R_0$ connected in series to the capacitor while the total charging time should then be $\tau = RC$ with $R=R_0+R_\mathrm{elec}$ and $C=1/(1/C_0+1/C_\mathrm{elec})$?}
%

\subsection{Co-axial cylindrical capacitor}
\begin{figure}
    \centering
    \includegraphics{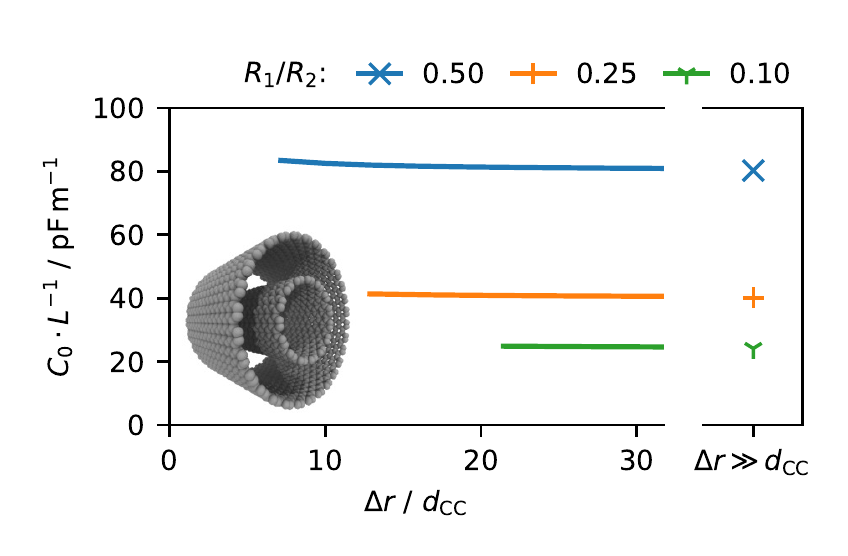}% 
    \caption{Dielectric capacitance of cylindrical capacitors for fixed ratios of $R_1/R_2$ with $R_1$ the inner and $R_2$ the outer tube radii.
    Results for co-axial carbon nanotubes computed with \acs{md} are indicated as lines.
    Marks denote analytical results for a structureless equivalent system.
    The difference between the radii $\Delta r=R_2-R_1$ is given relative to the characteristic bond length $d_\text{CC}$ between carbon atoms in graphene.}
    \label{fig:tube_in_tube}
\end{figure}

As a sanity check of our approach for systems which are periodic in just one dimension, we study the capacitance of two co-axial carbon nanotubes of radii $R_1$ and $R_2$, with $R_1<R_2$ (cf. inset of Fig.~\ref{fig:tube_in_tube}).
The vacuum capacitance $C_0$ of the co-axial carbon nanotubes can be calculated from the electrode-wise capacitance $\mathbf{\Tilde C}$\cite{Scalfi2020}.
At large radii, the atomic structure of the tubes should have a negligible effect, and thus the capacitance should approach that of structureless cylinders. 
The analytical line capacitance for a given ratio of the radii is $C_0/L = 2 \pi \epsilon_0 / \ln(R_2/R_1)$; in which $\varepsilon_0$ is the vacuum permittivity and $L$ is the length of the simulation box in the periodic dimension.
As shown in \reffig{fig:tube_in_tube} for various fixed ratios of the inner and outer tubes, the capacitance indeed converges to that of a structureless cylindrical capacitor when the radii are large compared to the bond length $d_\text{CC}$ between carbon atoms.
In the \ac{cpm} electrode atoms are assigned a Gaussian charge distribution
$\rho_i(\mathbf r) = q_i \left( {\eta^2} / {\pi} \right) ^ {3 / 2} \exp{\left[ - \eta ^ 2 \left( \mathbf r - \mathbf R_i \right) ^2 \right]}$
at their position $\mathbf R_i$ with the reciprocal charge width $\eta$.
In agreement with \citet{Serva2021} increased capacitances are observed for larger Gaussian width (i.e. smaller $\eta$) in \reffig{fig:tube_in_tube}.
However, the impact is almost negligible.

\acresetall
\section{Conclusions}
\label{sec:conclusions}

We presented the ELECTRODE package as an efficient implementation of the \ac{cpm} and closely related methods for the popular \acs{lammps} simulation environment.
Initially, the main goal was to bundle many different approaches to electrochemical simulations into one package and ensure that they are handled in the most computationally efficient way.
However, we also found interesting relationships between the two relevant charge-time contributions, i.e., water dipole relaxation and ion diffusion.
We also implemented several new features, such as the EW2D summation and a correction for systems periodic in just one dimension, whose capabilities and full potential have been scarcely explored and which also work independently of the \ac{cpm}.
Recent improvements to the \ac{cpm} such as the \ac{ff} method and a \ac{tf} model were included and compared for consistency to results found in the literature.
Remarkably, using the \ac{tf} model with varying \ac{tf} lengths has a suprising and complex impact on the water and ionic relaxation times.

These results demonstrate that the ELECTRODE package can efficiently simulate electrified interfaces, including unusual systems such as infinitely long charged nanotubes.
For a capacitor composed of co-axial carbon nanotubes, the vacuum capacitance agrees well within the limit of the analytical result of a structureless cylindrical capacitor and enables the investigation of curvature-dependent effects\cite{Seebeck2022ElucidatingSupercapacitors,Janssen2019CurvatureElectrodes} more rigorously in the future by avoiding interactions between the nanotubes through the periodic images.
Moreover, the charging process of a plate capacitor with an aqueous NaCl electrolyte in between illustrates vividly the differences between the range of methods introduced here and which are used to estimate the electrode charges.
Interestingly, in these simulations, it was observed that the water dipole initially responds very quickly to the applied potential but then slowly drops off as the ionic dipole slowly builds up, as if the water dipoles were shielded from the ions.

While the package is in a stable state, the development is ongoing and will include in the future features like a conjugate gradient solver or compatibility to TIP4P water models.
%

%TC:ignore
\section*{Supplementary Material}

The supplemental material provides more background on the \ac{tp}, boundary corrections and the \ac{ff} method.
Further, a description of the package interface to \acs{lammps} and an overview of the implemented classes are given.
More details on the simulations are provided, including plots of trajectories and of charging times as a function of the voltage, the \ac{tf} length and the time constant of the \ac{tp}.

\begin{acknowledgments}
Funded by the Deutsche Forschungsgemeinschaft (DFG, German Research Foundation) -- Projektnummer 192346071; 390794421 -- SFB 986 and GRK 2462.
\end{acknowledgments}

\section*{Data Availability Statement}

The ELECTRODE package has been merged into the official release of \acs{lammps} which is available under \url{github.com/lammps/lammps/tree/release}.

\bibliographystyle{jabbrv_apsrev4-1}
\bibliography{ludwig,robeme}
%TC:endignore

%% Acronyms
\begin{acronym}[ECU]
    \acro{p3m}[P\textsuperscript{3}M]{particle-particle particle-mesh}
    \acro{md}[MD]{molecular dynamics}
    \acro{cpm}[CPM]{constant potential method}
    \acro{ccm}[CCM]{constrained charge method}
    \acro{tp}[TP]{thermo-potentiostat}
    \acro{lammps}[LAMMPS]{the Large-scale Atomic/Molecular Massively Parallel Simulator}
    \acro{ff}[FF]{finite field}
    \acro{fd}[FD]{finite displacement}
    \acro{tf}[TF]{Thomas-Fermi}
    \acro{ew3d}[EW3D]{three-dimensional Ewald}
\end{acronym}

%TC:ignore
%\quickwordcount{main}
% \quickcharcount{main}
% \detailtexcount{main}
%TC:endignore

\makeatletter\@input{xx.tex}\makeatother

\end{document}

\title{SI - ELECTRODE: An electrochemistry package for atomistic simulations}
\author{Ludwig J. V. Ahrens-Iwers}
\affiliation{Institute of Advanced Ceramics, Hamburg University of Technology, Hamburg, Germany}%Lines break automatically or can be forced with \\

\author{Mathijs Janssen}%
\affiliation{Department of Mathematics, Mechanics Division, University of Oslo, N-0851 Oslo, Norway}%

\author{Shern R. Tee}%
\affiliation{Australian Institute for Bioengineering and Nanotechnology, The University of Queensland, Brisbane, Queensland, Australia}%

\author{Robert H. Mei{\ss}ner}
\affiliation{Institute of Polymers and Composites, Hamburg University of Technology, Hamburg, Germany}%
\affiliation{Helmholtz-Zentrum Hereon, Institute of Surface Science, Geesthacht, Germany}

%\email{s.tee@uq.edu.au, robert.meissner@tuhh.de}

\date{\today}% It is always \today, today,

% \acresetall

\maketitle
\section{Package overview}
\label{sec:package-overview}
\subsection{Interface}
The ELECTRODE package is designed as an add-on to \ac{lammps} and integrates seamlessly with most \ac{md} simulations carried out with that package. 
The user must first select one of two long-range Coulombic solvers (``kspace styles'' in \ac{lammps} terminology) included in ELECTRODE, which implement either an Ewald summation \cite{Ewald1921} (\texttt{ewald/electrode}) or a mesh-based summation \cite{Hockney1973, Eastwood1980} (\texttt{pppm/electrode}) providing additional methods to compute long-range contributions for the matrix $\mathbf A$ and vector $\mathbf b$ required in the \ac{cpm}.
The user can then include the following command in the simulation script:
%\begin{Verbatim}[fontsize=\small]
\begin{Verbatim}
fix <ID> <group> electrode/<mode> <value> <eta> 
\end{Verbatim}
followed by optional keywords.
Mandatory inputs are as follows:
\begin{itemize}
    \item \texttt{ID} is used for identification of the fix. The instantaneous applied potential for each electrode is available via \texttt{f\_<ID>[<x>]}.
    \item \texttt{group} specifies the \ac{lammps} group of particles to which the potential or charge \texttt{value} will be applied via \ac{cpm} or \ac{ccm}, respectively; additional groups could be specified with the \texttt{couple} keyword.
    \item \texttt{mode} chooses which of the following variants is performed:
    \begin{itemize}
        \item \texttt{conp}: constant potential
        \item \texttt{conq}: constrained charge
        \item \texttt{thermo}: \ac{tp} \cite{Deienbeck2021DielectricApproach}
    \end{itemize}
    \item \texttt{eta} specifies the reciprocal width of Gaussian smearing applied to the electrode charges.
\end{itemize}
The following optional keywords are available:
\begin{itemize}
    \item \texttt{couple <group-x> <value>} allows additional groups to be specified as an electrode. This keyword can be used multiple times to specify more than two electrodes.
    \item \texttt{etypes <types>} allows users to specify atom types exclusive to the electrode, so that \ac{lammps} can provide optimized neighborlists for faster calculations.
    \item \texttt{symm [on/off]} allows users to enforce charge neutrality for the electrodes via matrix pre-projection (cf. \refeq{M-eqn:S}).
    \item \texttt{ffield [on/off]} allows users to turn on or off the finite-field implementation \cite{Dufils2019SimulatingElectrode, Dufils2021ComputationalSimulations}. ELECTRODE automatically creates the electric field required to polarize the simulation box.
    \item \texttt{[write/read]\_[mat/inv] <path>} specifies a file path for input or output of the elastance or capacitance matrix.
    \item \texttt{temp <T-v> <tau-v> <seed>} specifies the parameters $T_v$, $\tau_v$ and the seed for the random number $X$ in \refeq{M-eqn:deissenbeck}. This keyword is only available with \texttt{mode} \texttt{thermo}, i.e. the \ac{tp}.
\end{itemize}
In addition, parameters for the \ac{tf} model can be set via:
\begin{Verbatim}
fix_modify <ID> tf <type> <length> <volume>
\end{Verbatim} 
with the parameters \texttt{length} and \texttt{volume} corresponding to the \ac{tf} length, $l_\mathrm{TF}$, and the reciprocal particle density, $d^{-1}$, as defined in \refcite{Scalfi2020ASimulations}, respectively.
Parameters need to be specified for each atom type in the electrode.
Further, the \texttt{pppm/electrode} long-range solver can be configured to calculate its part of the elastance matrix in two steps via:
\begin{verbatim}
kspace_modify amat [onestep/twostep]
\end{verbatim}
The \texttt{twostep} option can be faster for large electrodes and moderate mesh sizes but requires more memory\cite{Ahrens-Iwers2021}.

The atomic charges of the relevant electrode groups will then be dynamically updated over time, resulting in a \ac{md} simulation with \ac{cpm}, \ac{ccm} or \ac{tp} dynamics as selected. 
The computed electrode charges are accessible using standard \ac{lammps} commands. 

\subsection{Implementation}
\begin{figure}[tb]
    \centering
    \includegraphics{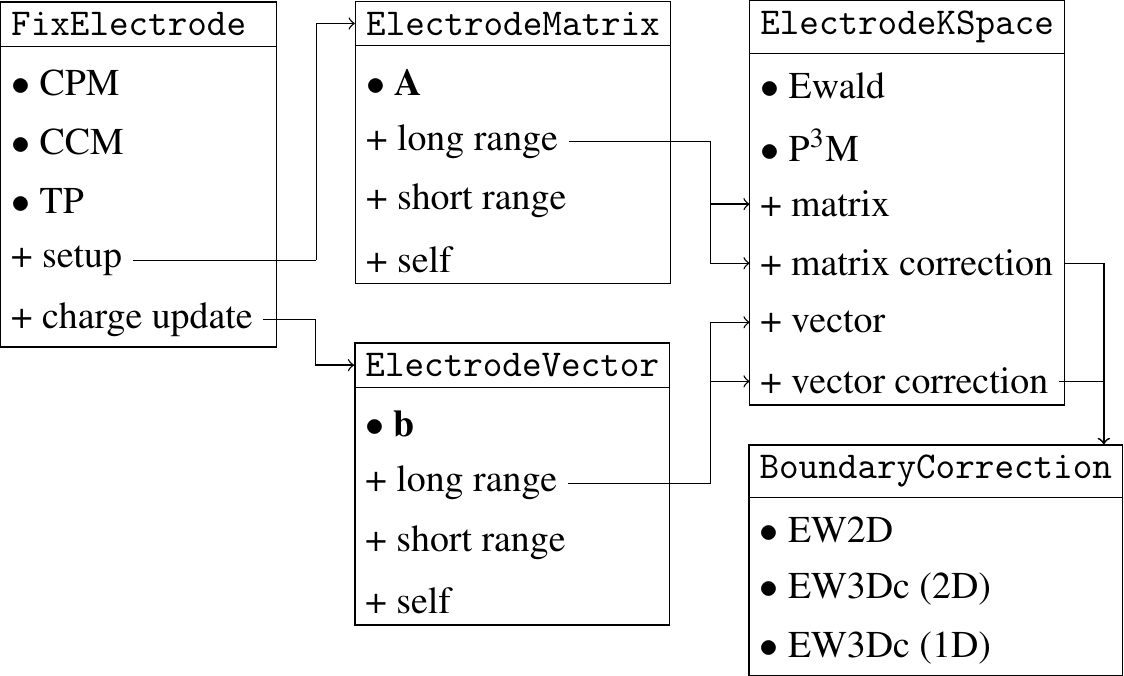}
    \caption{Overview of the package. 
        Each box represents a parent class. Implementations of the respective class are listed with bullet points. 
        Essential methods of each class are listed and their interactions are indicated with arrows.
        }
    \label{fig:overview}
\end{figure}
In contrast to previous versions, our implementation separates the class for charge updates from the computation of the elastance matrix and the electrolyte vector $\mathbf b$.
The \texttt{FixElectrode} class is derived from the \ac{lammps} \texttt{Fix} class and is used to determine and set electrode charges.
As shown in \reffig{fig:overview}, the computation of  the short-range, long-range and self interaction as well as the boundary correction are distributed across four additional classes. 

The \texttt{ElectrodeMatrix} and \texttt{ElectrodeVector} classes handle the computation of the elastance matrix and electrolyte vector, respectively, with each class containing its own code for the short-range Coulombic interactions.
For long-range interactions and boundary corrections, the two classes in turn delegate computations to implementations of the abstract \texttt{ElectrodeKSpace} class.
The two implementations inherit from the \ac{lammps} \texttt{KSpace} classes for the Ewald summation and \ac{p3m} method, respectively. 
For the functions of \texttt{ElectrodeKSpace}, additional Fourier transforms for the electrolyte charges and the Green's function at the position of electrode particles have been implemented \cite{Ahrens-Iwers2021}.
A \texttt{BoundaryCorrection} class was also added, which encapsulates the existing \ac{lammps} implementation of the \acs{ew3d}c method \cite{Smith1981ElectrostaticCrystals, Yeh1999} and adds new implementations for a 2D Ewald summation \cite{Parry1975TheCrystal, Hu2014InfiniteInterfaces} and a correction intended for 1D-periodic systems \cite{Brodka2004Three-dimensionalDirection, Smith1981ElectrostaticCrystals}.

As developers, we have found that an object-oriented programming approach has offered significant advantages during our work in writing the ELECTRODE package. 
% %
Splitting the code across distinct classes has significantly clarified information flows between the different parts of the ELECTRODE package, leading to easier maintenance and optimization.

\subsection{Comparison with other implementations}
\label{sec:metalwalls}
Another open source implementation of the \ac{cpm} is found in \textit{MetalWalls}.\cite{Marin-Lafleche2020}
In contrast to ELECTRODE it is a stand-alone \ac{md} code and is dedicated to electrochemical systems.
\textit{MetalWalls} offers features such as the conjugate gradient and the mass-zero\cite{Coretti2020Mass-zeroSystems} method for the energy minimization, dipole interactions and the TIP4P water model, which are not yet available in ELECTRODE.
Further, \textit{MetalWalls} offers tools for post processing that are particularly useful for electrochemical simulations.
Fortunately, many of these scripts can also be used for \ac{lammps} trajectories.

On the other hand, ELECTRODE is integrated into \ac{lammps} which supports a wide range of force fields and \ac{md} features which may not be available in \textit{MetalWalls}, including many-body and machine-learning interatomic potentials, interfaces with quantum calculation software, multi-replica simulations and non-equilibrium \ac{md}.
ELECTRODE uses self-contained Coulombic solvers to determine electric potentials, including requesting an appropriate neighbor list from \ac{lammps}.
Thus, ELECTRODE  will work with any force field that includes long-range Coulombic interactions, including force field combinations enabled under the \ac{lammps} keyword `hybrid'.
(The usual caveat applies: just because an \ac{md} simulation can run does not ensure it is physically valid.)
ELECTRODE reuses many data structures native to \ac{lammps} (such as neighbor lists for nearby interactions and \ac{p3m} implementations), thus benefiting from the high scaling efficiency that has been documented in \ac{lammps} publications.
Additionally, ELECTRODE can use the efficient \ac{p3m} method for long-range interactions required for a \ac{cpm}.\cite{Ahrens-Iwers2021}

A notable omission from ELECTRODE as compared to \textit{MetalWalls} is that the charge update of ELECTRODE is not synchronized with other features which attempt to minimize energy mid-step, such as Drude model polarizable force-fields and charge equilibration of mobile particles in ReaxFF and other reactive force-fields.
\textit{MetalWalls} includes an implementation of polarizable ion models represented by point dipoles, where electrode charges and point dipoles are simultaneously updated in a synchronized energy minimization (using conjugate gradients).
By contrast, in an ELECTRODE simulation with (for example) Drude dipoles, electrode charges will not be re-optimized after Drude dipoles have been optimized (or vice versa), so in theory electrode charges may be out of sync with Drude dipole configurations, and simulation results will change based on the order in which commands are entered in the script.
In practice, stable integration requires that both electrode charges and Drude dipoles do not change abruptly over the integration timestep.
We therefore think that the absence of synchronized minimization is not a major obstacle to using ELECTRODE with polarizable force fields, but look forward to verifying this in future with detailed simulations.

Another available implementation in \ac{lammps} is the `USER-CONP' code by \citet{Wang2014EvaluationCapacitors}
This code appears to be no longer maintained and is poorly optimized compared to other implementations. A follow-on version entitled `USER-CONP2'\cite{Tee_2022} was written and maintained by one author of this paper, and most of its features have now been incorporated into the ELECTRODE package.
An implementation of \ac{cpm} \ac{md} for the OpenMM software \cite{Eastman2013} can also be found at \url{https://github.com/jmcdaniel43/Fixed_Voltage_OpenMM}, and may be of interest to users with access to graphical processing units (GPUs) for computational acceleration.

\section{Simulation details}

A capacitor with an aqueous NaCl electrolyte between two gold electrodes has been adapted from an example provided in the
\textit{MetalWalls} repository\cite{Marin-Lafleche2020}.
For a consistent comparison, systems are equilibrated with the respective method at 0\,V for the \ac{cpm} and \ac{tp} methods and at 0\,e$^-$ for the \ac{ccm} methods.
After an equilibration time of 5\,ns, a total number of 100 equilibrated structures are written at an interval of 10\,ps.
For each of the resulting structures a charging process is modeled by applying a potential difference of 2\,V, %see \reffig{fig:bi-exp} for the evolution of the total z-dipole component $M_z$ during such a charging process, 
or an equivalent of roughly 4.4\,e$^-$, depending on the method.
\begin{figure}[tbp]
    \centering
    \includegraphics{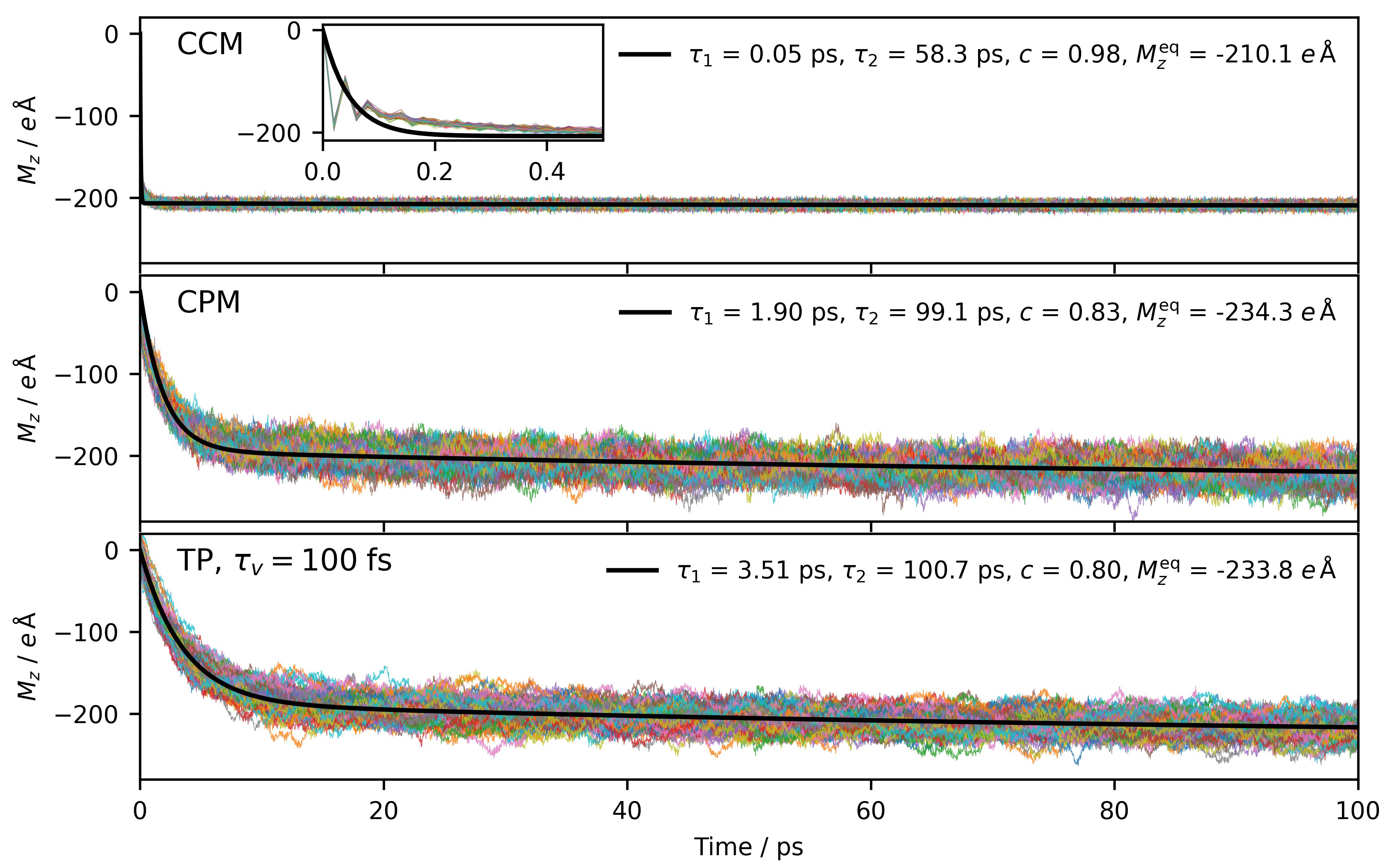}
    \caption{Bi-exponential function fit to 100 charging trajectories of the respective method.}
    \label{fig:bi-exp}
\end{figure}
Charging times $\tau_1$ and $\tau_2$ are obtained by fitting a bi-exponential charging function
\begin{equation}
    M_z(t) = M_z^\mathrm{eq} \left[1 - c \exp(-t / \tau_1) - (1-c)\exp(-t / \tau_2)\right]
    \label{eqn:bi-exp}
\end{equation}
to the $z$-component of the average electrolyte dipole of the 100 uncorrelated and individual runs.
Examples of the temporal evolution and the fitting are given in \reffig{fig:bi-exp} for the \ac{ccm}, \ac{cpm} and the \ac{tp}.
A summary of the fit parameters for the various charging runs with different methods is given in \reftab{tab:fits}.
\begin{table}[tbp]
\begin{tabular}{ l|r|r|r|r}
  Method & \multicolumn{1}{c|}{$\tau_1$ /  ps} & \multicolumn{1}{c|}{$\tau_2$ /  ps} & \multicolumn{1}{c|}{$c$} & \multicolumn{1}{c}{$M_z^{\mathrm{eq}}$ /  $e\,\text{\AA}$} \\
  \hline
  CCM uniform & 0.04 $\pm$ 0.00 & 53.5 $\pm$ 2.8 & 0.98 $\pm$ 0.00 & -200.4 $\pm$ 0.0 \\
  CCM & 0.05 $\pm$ 0.00 & 58.3 $\pm$ 3.1 & 0.98 $\pm$ 0.00 & -210.1 $\pm$ 0.0 \\
  CPM uniform & 2.00 $\pm$ 0.03 & 93.4 $\pm$ 1.3 & 0.81 $\pm$ 0.00 & -241.1 $\pm$ 0.1 \\
  CPM & 1.90 $\pm$ 0.03 & 99.1 $\pm$ 1.5 & 0.83 $\pm$ 0.00 & -234.3 $\pm$ 0.1 \\
  TP uniform & 3.49 $\pm$ 0.04 & 93.4 $\pm$ 1.3 & 0.79 $\pm$ 0.00 & -241.2 $\pm$ 0.1 \\
  TP, $\tau_v = 500\,\text{fs}$ & 9.60 $\pm$ 0.09 & 107.2 $\pm$ 1.7 & 0.79 $\pm$ 0.00 & -234.3 $\pm$ 0.1 \\
  TP, $\tau_v = 100\,\text{fs}$ & 3.51 $\pm$ 0.04 & 100.7 $\pm$ 1.4 & 0.80 $\pm$ 0.00 & -233.8 $\pm$ 0.1 \\
  TP, $\tau_v = 20\,\text{fs}$ & 2.12 $\pm$ 0.03 & 94.1 $\pm$ 1.4 & 0.82 $\pm$ 0.00 & -234.1 $\pm$ 0.1 \\
  CPM, $l_{\mathrm{TF}} = 1.5\,\text{\AA}$ & 1.40 $\pm$ 0.03 & 103.0 $\pm$ 2.2 & 0.85 $\pm$ 0.00 & -153.7 $\pm$ 0.1 \\
  CPM, $l_{\mathrm{TF}} = 1.0\,\text{\AA}$ & 1.64 $\pm$ 0.03 & 94.2 $\pm$ 2.0 & 0.86 $\pm$ 0.00 & -184.0 $\pm$ 0.1 \\
  CPM, $l_{\mathrm{TF}} = 0.5\,\text{\AA}$ & 1.82 $\pm$ 0.03 & 89.9 $\pm$ 1.4 & 0.82 $\pm$ 0.00 & -216.8 $\pm$ 0.1 \\  
  CPM 1.0 V & 1.87 $\pm$ 0.05 & 90.0 $\pm$ 3.1 & 0.84 $\pm$ 0.00 & -115.0 $\pm$ 0.1 \\
  CPM 0.5 V & 1.73 $\pm$ 0.10 & 63.9 $\pm$ 5.3 & 0.85 $\pm$ 0.01 & -57.3 $\pm$ 0.1 \\
  CPM 0.3 V & 2.04 $\pm$ 0.18 & 50.3 $\pm$ 10.2 & 0.88 $\pm$ 0.02 & -34.4 $\pm$ 0.1 \\
\end{tabular}
\caption{Parameters for the bi-exponential fit using \refeq{eqn:bi-exp} with residual standard deviations. \ac{cpm} simulations without specification of voltage are at 2.0\,V.}
\label{tab:fits}
\end{table}
\begin{figure}
    \centering
    \includegraphics{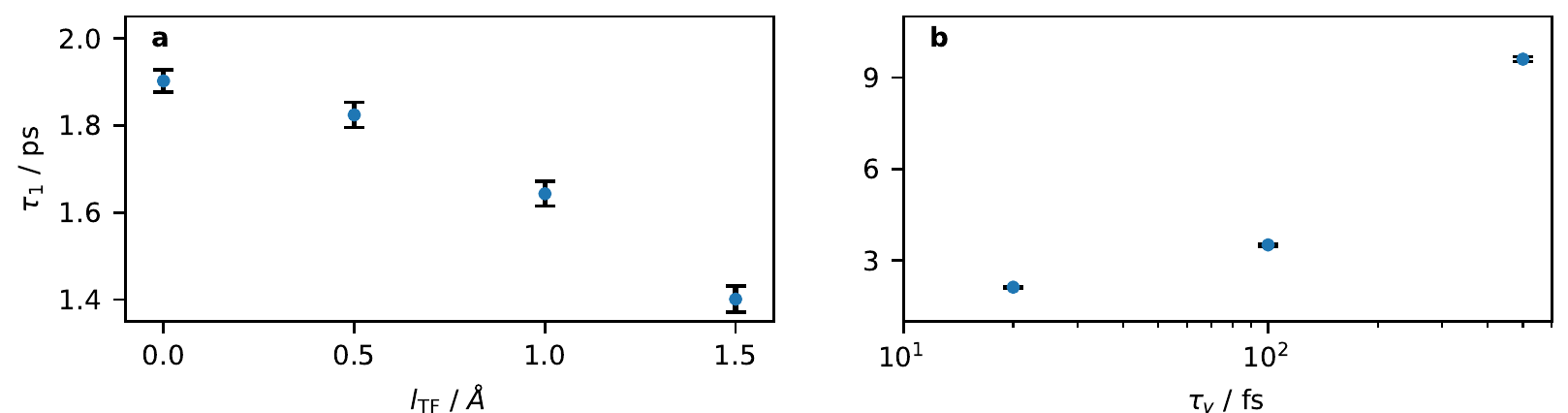}
    \caption{%
            Water dipole relaxation time constants $\tau_1$ and standard deviations for (a) the \ac{cpm} with varying Thomas-Fermi length $l_\mathrm{TF}$ and (b) the \ac{tp} with different \ac{tp} time constants $\tau_v$.
            $l_\mathrm{TF} = 0$ is identical with the standard \ac{cpm} in \reffig{M-fig:tau}.
            }
    \label{fig:tau-tf-tps}
\end{figure}
\reffig{fig:tau-tf-tps} features a \ac{cpm} setup with the \ac{tf} approach and various $l_\mathrm{TF}$. 
While not affecting $\tau_2$, increasing $l_\mathrm{TF}$ leads to a faster water dipole reorientation (cf. Fig.~\ref{fig:tau-tf-tps}a) and an up to 35\% smaller equilibrium dipole compared to a pure \ac{cpm} with $l_\mathrm{TF}=0$\,{\AA} due to the screening of the charge in the metal electrode.
In \reffig{fig:tau-tf-tps}b different potentiostat couplings for the \ac{tp} are investigated.
It is expected that the \ac{tp} alters charging times because it includes a dissipation term which leads to a delayed response in terms of a weaker coupling to the external (electron) reservoir via $\tau_v=R_0C_0$.\cite{Deienbeck2021DielectricApproach}
Increasing $\tau_v$ leads thus to larger charging constants for the water dipole reorientation $\tau_1$, but does not affect the ionic diffusion relaxation time constant $\tau_2$ (cf.~\reftab{tab:fits}).
%
% The \ac{pnp} equations\cite{Bazant2004diffuse} suggest an ionic time constant of $\tau_2=0.5$\,ps, as derived in the main text.
% %
% The discrepancy between our MD results and the \ac{pnp} model may be related to the nanoconfinement\cite{Deienbeck2021DielectricApproach}, since the dielectric constant of the electrolyte depends strongly on it, \textit{and} the finite salt concentration.
% %
% While it has been observed that the ionic $RC$-time resulting from the \ac{pnp} equations agrees in \refcite{Ma2022dynamics} using DDFT calculations up to 1\,V, we would expect better agreement at lower voltages. 
% %
% The opposite was observed in \reffig{fig:voltage-tau}, i.e., the ion relaxation time $\tau_2$ decreases for lower potentials.
% %
% However, some dependence of the applied potential $\Delta v_0$ on $\tau_2$ is not unusual.
% %
% In the simplest case, one replaces the Helmholtz capacitance (which gives \refeq{M-eq:RC}) with the Gouy-Chapman capacitance, which includes an additional $\cosh(\Delta v_0)$ term; therefore, $\tau_2$ increases with $\Delta v_0$.
% %

\begin{figure}[tbp]
    \centering
    \includegraphics{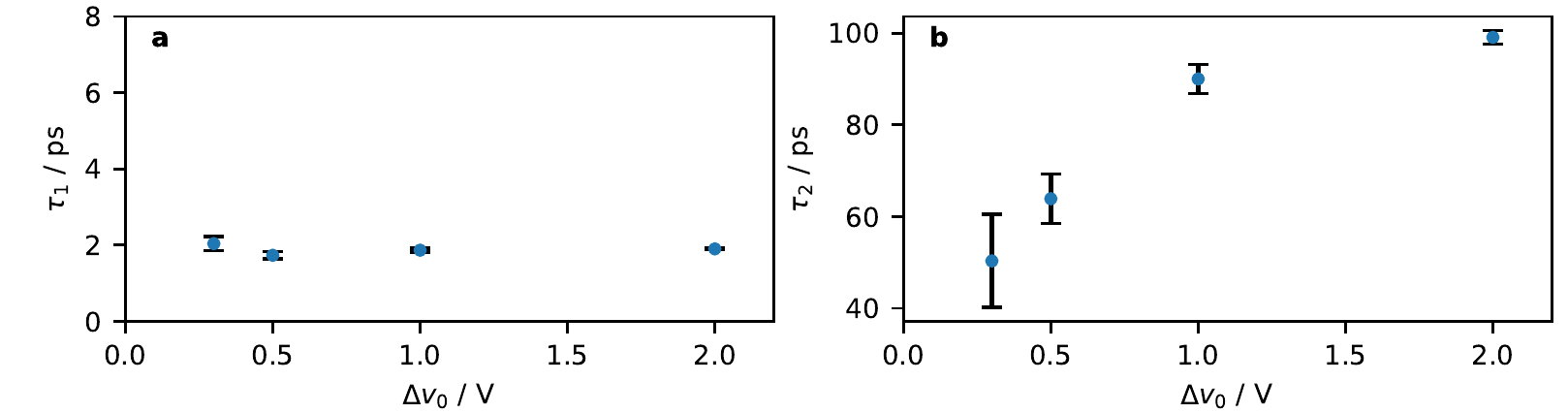}
    \caption{$\tau_1$ and $\tau_2$ as function of voltage $v_0$ applied with the \ac{cpm}. Fitted values and residual standard deviations for fitted relaxation times are obtained from averaging over 100 uncorrelated runs.}
    \label{fig:voltage-tau}
\end{figure}

\refFig{fig:voltage-tau} illustrates the dependence of the voltage $v_0$ on the two relaxation times $\tau_1$ and $\tau_2$.
Interestingly, the voltage here only slightly affects $\tau_1$, but strongly changes the ionic relaxation time $\tau_2$ at voltages below 1\,V.
\FloatBarrier

\bibliographystyle{apsrev4-1} % TODO
\bibliography{robeme, ludwig}

%% Acronyms
\begin{acronym}[ECU]
    \acro{p3m}[P\textsuperscript{3}M]{particle-particle particle-mesh}
    \acro{md}[MD]{molecular dynamics}
    \acro{cpm}[CPM]{constant potential method}
    \acro{ccm}[CCM]{constrained charge method}
    \acro{tp}[TP]{thermo-potentiostat}
    \acro{lammps}[LAMMPS]{the Large-scale Atomic/Molecular Massively Parallel Simulator}
    \acro{ff}[FF]{finite field}
    \acro{fd}[FD]{finite displacement}
    \acro{tf}[TF]{Thomas-Fermi}
    \acro{ew3d}[EW3D]{three-dimensional Ewald}
    \acro{dwcnt}[DWCNT]{double walled carbon nanotube}
    \acro{pnp}[PNP]{Poisson-Nernst-Planck}
\end{acronym}

\makeatletter\@input{yy.tex}\makeatother